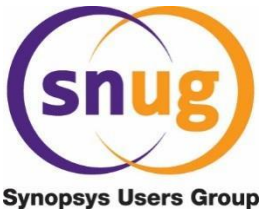

# Multiple-lithography-compliant verification for standard cell library development flow

Yongfu Li, Wan Chia Ang, Chin Hui Lee,
Kok Peng Chua, Yoong Seang Jonathan Ong, Chiu Wing Colin Hui

GLOBALFOUNDRIES

**ABSTRACT**

Starting from 22-nm, a standard cell must be designed to be full lithography-compliant, which includes Design Rule Check, Design-for-Manufacturability and Double-Patterning compliant. It has become a great challenge for physical layout designers to provide a full lithography-compliant standard cell layout that is optimized for area, power, timing, signal integrity, and yield. This challenge is further exacerbated with abutted single- and multiple-height standard cells. At present, different foundries and library vendors have different approaches for full lithography-compliant library preparation and validation. To the best of our knowledge, there is no single tool integrates all types of lithography-compliant check in standard cell libraries validation flow. In this work, we will demonstrate multiple lithography-compliant verifications for standard cell library development flow. Validation flow and detailed algorithm implementation will be explained to assist engineers to achieve full lithography-compliant standard cell libraries. An area-efficient standard cell placement methodology will also be discussed to validate the issues arises from standard cell abutment.

## Table of Figures

## Table of Tables

## I. Introduction

With the continuous scaling on the CMOS technology, the new technology is facing the bottleneck using 193i optical lithography process. Pattern-related defects continue to increase and limit the number of good die per wafer. The classical rule-based Design Rule Check (DRC) approach is no longer sufficient to guarantee 100% pattern printability. Design-for-Manufacturability (DFM) compliance checking is required to identify manufacturing weak-points and prevent catastrophic errors such as open (necking) and shorts (bridging) issues. Starting from the GLOBALFOUNDRIES 22-FDX technology, all the designs are required to be color-compliant due to the Double-Patterning technology (DPT). However, it is hard to achieve full lithography-compliant standard cell Intellectual Property (IP) libraries due to the sheer number of layout combinations arises from the different single- and multiple-height standard cells abutment conditions. Each of these has challenging implications prohibits the design to achieve optimal performance, power and area (PPA). Therefore, early assessment of the IP design constraints imposed by the technology is necessary to ensure a full lithography-compliant design.

DFM techniques, such as lithographic process variability simulation, chemical mechanical polishing (CMP) simulation, and critical area analysis (CAA) simulation, provide many opportunities to improve layout and thus further enhance the overall yield. These DFM techniques help to bridge the gap between design and manufacturing, enabling early ramp to good yield [1], [2]. For example, in a chip design, there are layout patterns that are either difficult to print during lithography, or very susceptible to process variations. These patterns are commonly referred to lithography hotspot and it is required to identify and fix early before the design goes into the production phase. Therefore, a combination of DRC and lithography simulation is required to check the entire design library to minimize the systematic yield-loss. GLOBALFOUNDRIES offers a quick-lithography verification solution, named DRC+, which is based on pattern matching technology, can rapidly scan for fab-learned critical lithography patterns in a design and thus detects lithographic hotspots very effectively such that it is scalable to large chip verification [3].

In the DPT, each design layer is partition into two separate masks (indicated by different colors on layout) to enable better printability. In the chip design phase, the color violation usually occurs at the cell boundaries mainly because of insufficient distance of the metal layers from the boundary. In the current design methodology, color spacing is checked during the placement stage to prevent DP violations due to standard cell abutment [4]. However, this methodology does not guarantee optimal chip design.

To achieve optimal chip design, all the standard cells must achieve full lithography-compliant during the IP qualification stage. However, with the number of standard cells increases with various design variants and functionality, the total number of cell placement combinations grows exponentially. More combinations yield more physical area and polygons which impacts the verification run time. Therefore, it is important to provide an integrated verification tool to assist layout engineer to identify lithography weak-points at the standard cell abutment boundaries so that they are able to address the errors.

In this work, we make the following contributions.
1. We proposed a simplified abutment unit testcase to eliminates redundant cases and maintains 100% coverage of all standard cell abutment topologies with minimum area.
2. We provide an integrated utility tool based on Synopsys ICC and ICV platform for ICV DRC+ and ICV DRC verifications.
3. We demonstrated through the use of the case study with 14nm standard cells IP.

The rest of this paper is organized as follows. Section II details the implementation of our Tcl procedure script. The verification result is discussed in the Section III and conclusions are given in Section IV.

## II. Design Implementation

The implementation of multiple lithography-compliant standard cell library validation flow is based on a customized Tcl procedure script, which leverages on the Synopsys IC Compiler commands [5]. This is a required tool to help layout engineers or IP engineers in checking and fixing potential lithography weak-points at the boundary of the standard cell abutment, especially during the library development phase.

In this work, we aim to simplify the use-model for the validation flow. As such, this Tcl procedure only requires user to provide the list of milkyway libraries as input. The simple-to-use approach reduces the barriers to adopting of new tool, especially our target users who might not be well-versed with programming and digital design flow.

The associated IC commands to implement the flow is summarized in Figure 1 and the customized Tcl procedure can be broken down into the following phases.
   A.  Standard cell profiling
   B.  Reduction techniques for standard cell abutment placement
   C.  Verilog and DEF files generation
   D.  Design Rule Verification (Include color-decomposition verification) and DRC+ Verification.

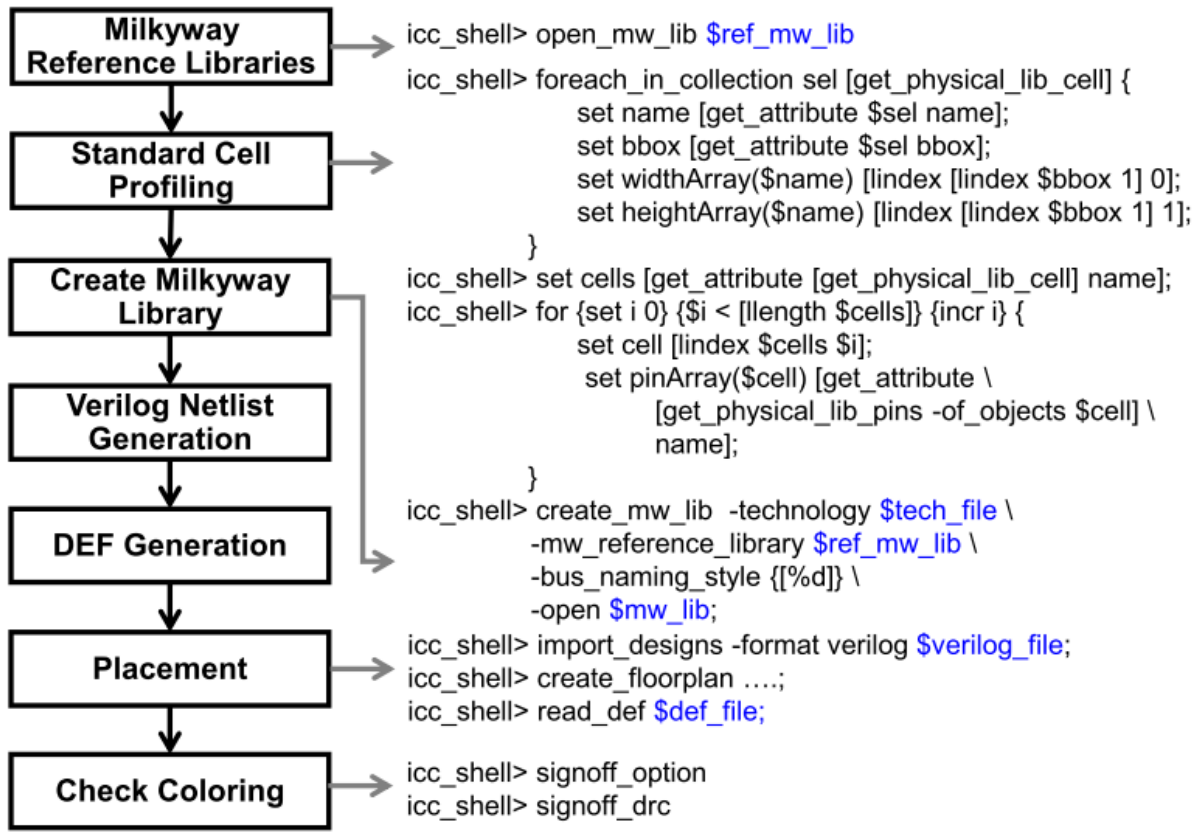

Figure 1. The proposed multiple full lithography-compliance standard cell validation flow.

SNUG 2017SNUG 2017

## A. Standard Cell Profiling

To enumerate all the standard cells abutment conditions, we need the basic standard cell information from the list of milkyway libraries. This can be done with the following algorithm:

```tcl
array set widthArray {}; # Initial the array
array set heightArray {}; # Initial the array
array set pinArray {}; # Initial the array

foreach lib $mw_libs {
    open_mw_lib $lib; # Open the milkyway database;
    foreach_in_collection sel [get_physical_lib_cell] {
        # The standard cell physical information is stored in the Tcl array.
        set name [get_attribute $sel name]; # Get the cell name;
        set bbox [get_attribute $sel bbox]; # Get the cell boundary information;
        set widthArray($lib,$name) [lindex [lindex $bbox 1] 0]; # Get the cell width
        set heightArray($lib,$name) [lindex [lindex $bbox 1] 1]; # Get the cell height
    }
    # The standard cell pin information is stored in the Tcl array.
    set cells [get_attribute [get_physical_lib_cell] name];
    for {set i 0} {$i < [llength $cells]} {incr i} {
        set cell [lindex $cells $i];
        set pinArray($lib,$cell) [get_attributes [get_physical_lib_pins –of_objects $cell] name];
    }
}
```

The algorithm translates the standard cell physical information into Tcl arrays, which will be used for the Verilog netlist and design exchange format (DEF) file generations. Figure 2 summarizes the standard cells physical information in histogram plots for two libraries, one with single-height and another with multiple-height.

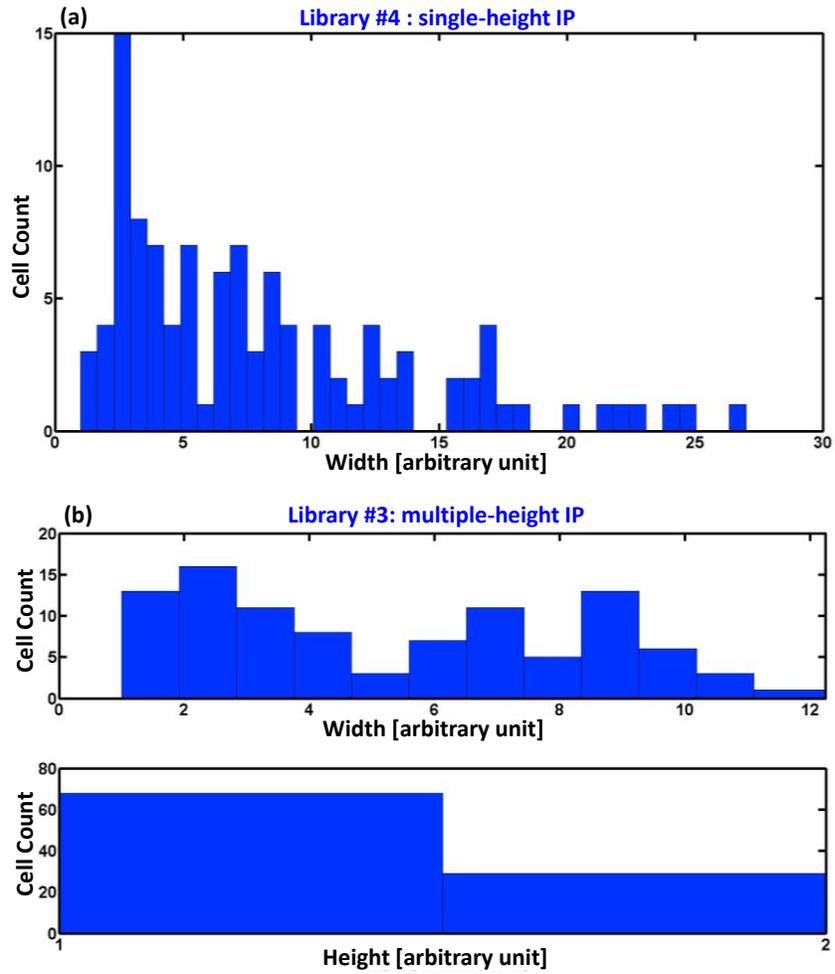

Figure 2: The standard cells physical information in histogram plot for (a) single-height and (b) multiple-height IPs.

## B. Reduction techniques for Standard Cell Abutment Placement

In the chip design, there are four possible orientations of a standard cell that can be placed in the layout, as shown in Figure 3. The R0- and R180-orientated cells are the default cell orientation and cell rotated by 180 degree. The MX- and MY-orientated cells are cells mirrored along the X-axis and Y-axis, respectively. A conventional single-height standard cell is designed with the power signal (VDD) and the ground signal (VSS) at the top and bottom rails, respectively. Therefore, the single-height standard cells with R0- and MY-orientations are legally placed at the same row while the single-height standard cells with MX- and R180-orientations are placed at the alternate row. This arrangement allows the standard cells to be placed along pre-defined, interleaved horizontal VDD and VSS rails during placement phase. As shown in Figure 4, a standard cell (highlighted in 'grey') can be placed in the layout with orientation of R0 and be surrounded by other cells with orientation of R0 or MY.

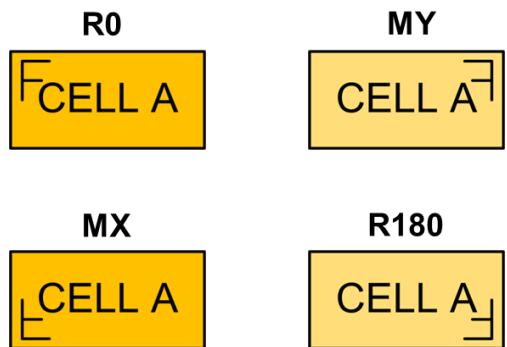

Figure 3. Four possible legal orientations of standard cell placement in the chip design.

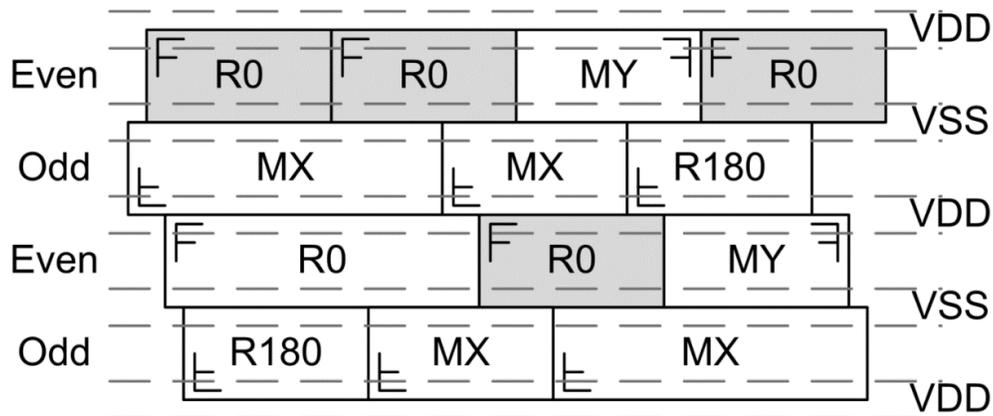

Figure 4. Standard cells placement in the chip design.

To reduce the number of permutation, we need to consider the following scenarios:
- Single-height type A-A: Single-height side-to-side standard cell abutment with identical cells.
- Single-height type A-B: Single-height side-to-side standard cell abutment with different cells.

As shown in Figure 5, there are a total of four possible permutation pairs of placement for single-height type A-A condition. The number of standard cells can be reduced by overlapping and placing the entire permutated pairs in a single row. As shown in Figure 6, the R0-R0 permutation pair overlaps with R0-MY and MY-R0 permutation pairs on the right and left sides, respectively. The MY-MY permutation pair is discarded in the proposed abutment placement because it is a mirror version of R0-R0. The proposed abutment placement can achieve an area reduction of 2 times.

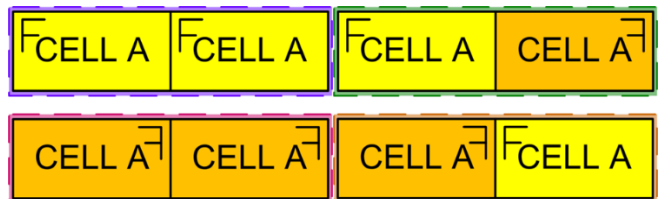

Figure 5. Four possible permutation pairs of placement for single-height type A-A condition.

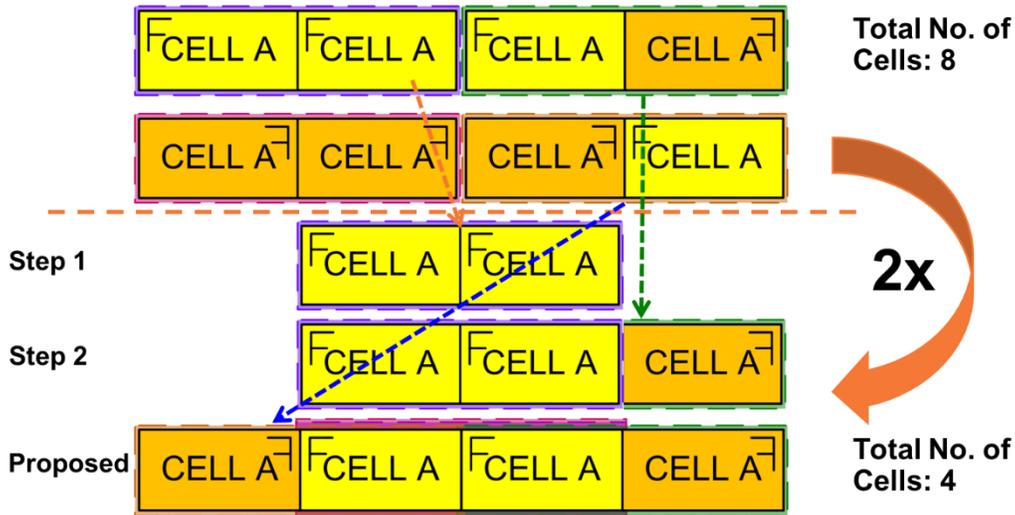

Figure 6. The proposed single-height type A-A cell abutment placement.

For single-height type A-B condition, there are a total of eight possible permutation pairs of placement, as shown in Figure 7. As illustrated in Figure 8, using the similar technique described earlier, the number of standard cells can be reduced by overlapping and placing the entire permutated pairs in a single row. The proposed abutment placement can achieve an area reduction of 3.2 times. Furthermore, there is a 2 times reduction in the number of standard cells because single-height type A-B condition is the same as single-height type B-A condition.

For a standard cell library with *N* number of cells, the total number permutation is reduced from $8*N + 8*(N-1)*N$ to $4*N + 2.5*(N-1)*N$. For example, in the 14-nm technology node, the number of single-height standard cells in a library is approximately 1000 cells, which translate to a total of 8,000,000 cells to achieve 100% abutment coverage using conventional placement method. Our proposed abutment placement requires only 2,501,500 cells, with about 3.2 times reduction. In this work, seven libraries were tested and the cells counts of the generated testcases match with our theoretical calculation. The results are listed in Table 1.

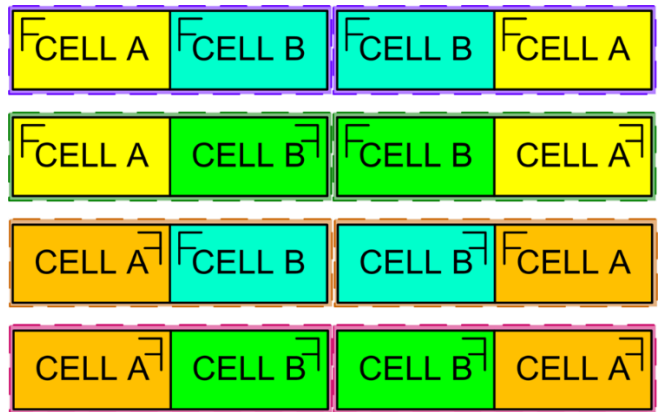

Figure 7. Eight possible permutation pairs of placement for single-height type A-B condition.

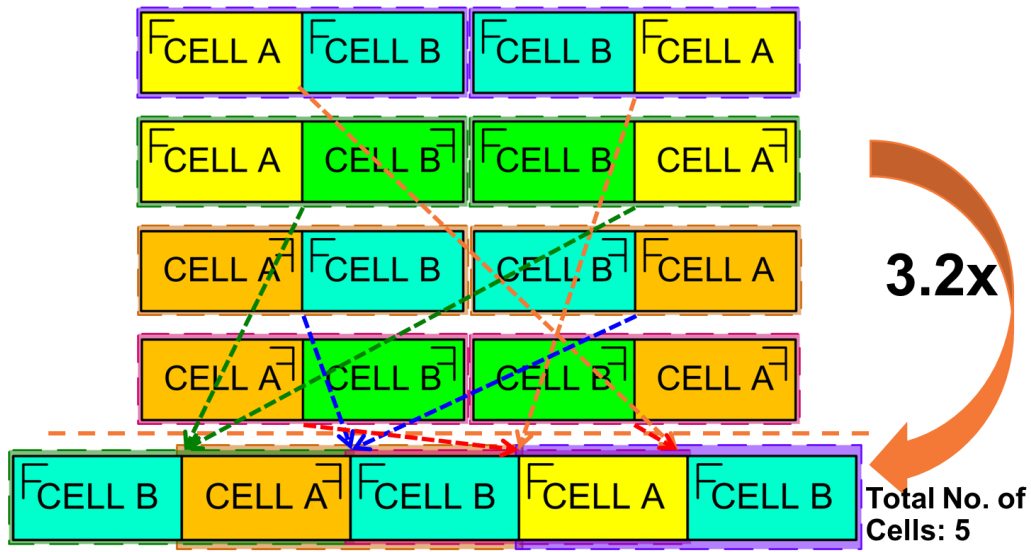

Figure 8. Proposed single-height type A-B cell abutment placement.

In addition to single-height cells abutment, we have considered multiple-height cells in the standard cell library. Multiple-height cells such as level shifter cell are designed with an integer multiple of the single-height standard cells. Therefore, we have to consider two additional placement conditions, which are single-and-multiple-height placement and multiple-multiple-height placement. The single-height Type A-A can be applicable to multiple-height Type A-A placement. For single-height and multiple-height cell placement, we extended the proposed abutment placement from single-height Type A-B to single-multiple-height Type A-B placement, as shown in Figure 9. Without the loss of generality, this work did not consider multiple-height-multiple-height Type A-B, however, the methodologies can be extended from the single-multiple-height Type A-B abutment placement.

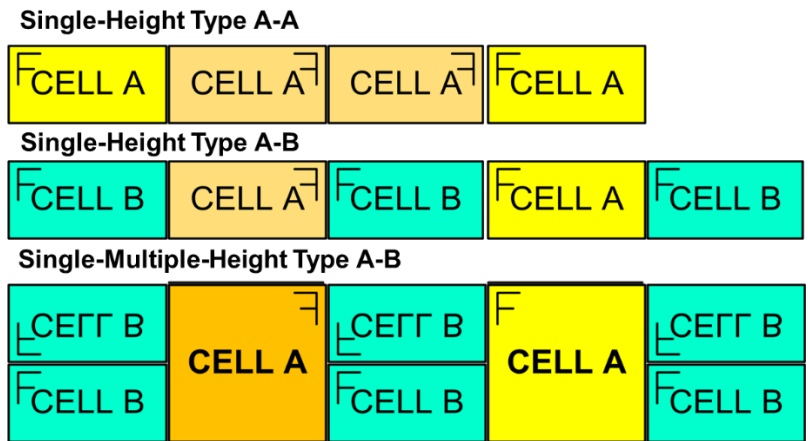

Figure 9. The proposed area-efficient cells abutment placement.

Table 1: Cells count in the proposed standard cells abutment placement.

| Libraries | Number of Tracks | Single-height Cells Count | Multiple-height Cells Count | Total Cells Count, $N$ | Cells Count in Testcase, [$4*N + 2.5*(N-1)*N$ for single-height cells only] |
|---|---|---|---|---|---|
| 1 | 9 | 470 | - | 470 | 552959 |
| 2 | 9 | 470 | - | 470 | 552959 |
| 3 | 9 | 68 | 21 | 89 | 24220 |
| 4 | 9 | 108 | - | 108 | 29322 |
| 5 | 10.5 | 470 | - | 470 | 552959 |
| 6 | 10.5 | 41 | - | 41 | 4264 |
| 7 | 10.5 | 108 | - | 108 | 29322 |

## C. Verilog netlist and DEF file generation.

To implement the proposed area-efficient abutment placements in the Synopsys IC Compiler, we have adopted the netlist and DEF placement flow.

The Verilog netlist describes the logic cell required the cell abutment and all the different types of abutment placement can be generalized in the illustration shown in the Figure 10. For example, for the single-height type A-A placement, it requires four identical cell, which is described under the module scell_<typeA> Verilog netlist.

The DEF file describes the physical layout of the cell abutment. It requires the basic standard cell information such as width and height, which is stored in the Tcl arrays during the standard cell profiling phase. The orientation of each cell is derived based on the type of abutment placement. For example, as shown in Figure 11, for the single-height type A-A placement, the cell orientation is N – FN – FN – N.

The Tcl procedure generates the Verilog netlist and DEF files according to the different types of abutment placement. The top level Verilog module includes all the abutment permutations to ensure 100% coverage of standard-cell placement topologies.

**Single-Height Type A-A**
**Multiple-Height Type A-A**
```
module scell_<typeA> ();
    <typeA> U1 (.*);
    <typeA> U2 (.*);
    <typeA> U3 (.*);
    <typeA> U4 (.*);
endmodule
```

**Single-Height Type A-B**
```
module scell_<typeA>_<typeB> ();
    <typeB> U1 (.*);
    <typeA> U2 (.*);
    <typeB> U3 (.*);
    <typeA> U4 (.*);
    <typeB> U5 (.*);
endmodule
```

**Single-Multiple-Height Type A-B**
```
module mcell_<typeA>_<typeB> ();
    <typeB> U1 (.*);
    <typeB> U2 (.*);
    <typeA> U3 (.*);
    <typeB> U4 (.*);
    <typeB> U5 (.*);
    <typeA> U6 (.*);
    <typeB> U7 (.*);
    <typeB> U8 (.*);
endmodule
```

**Top Module**
```
module TOP ();
    scell_<typeA> U1 ();
    scell_<typeA>_<typeB> U1 ();
    mcell_<typeA>_<typeB> U2 ();
endmodule
```

Figure 10. Verilog netlist for different standard cell abutment.

SNUG 2017

```
Example of the generated DEF file

VERSION 5.6 ;
DESIGN TOP;
TECHNOLOGY ROUTE ;
UNITS DISTANCE MICRONS 1000 ;
COMPONENTS 4;
 - sinst_<typeA>\U1 <TYPEA> + PLACED ( 0 0 ) N ;
 - sinst_<typeA>\U2 <TYPEA> + PLACED ( 200 0 ) FN ;
 - sinst_<typeA>\U3 <TYPEA> + PLACED ( 400 0 ) FN ;
 - sinst_<typeA>\U4 <TYPEA> + PLACED ( 400 0 ) N ;
END COMPONENTS
DIEAREA ( 0 0 ) ( 600 0 ) ;
END DESIGN
```

Figure 11. DEF file to define the placement of the standard cells abutment.

The following IC Compiler commands detail the floorplan implementation of the abutment placement:

   create_mw_lib \
      -technology $tech_file \
      -mw_reference_libraray $mw_libs \
      -bus_naming_style {[%d]} \
      -open $lib; # Create the milkyway library
   import_design –format Verilog $verilog_file; # Read in the Verilog file
   create_floorplan …; # Create the floorplan (Optional)
   read_def $def_file; # Read in the DEF file

## D. Design Rule Check (DRC) and DRC+ Verifications.

After the placement of the standard cells in the IC Compiler LayoutWindow, our utility provides the following physical verifications and summarizes the errors at the abutted cells' boundaries.
- Design Rule Check (DRC)
- GLOBALFOUNDRIES Pattern matching (DRC+)

The IC Validator In-Design feature in IC Compiler provides the ability to use the IC Validator tool to perform the DRC and DRC+ physical verifications [6].

The following IC Compiler commands detail the IC Validator setup and perform physical verifications in IC Compiler:

```
# DRC Physical Verification
if {[file exists $drc_runset]} {
    set_physical_signoff_options –exec_cmd {icv} –drc_runset {$drc_runset}; # DRC Setup
    report_physical_signoff_options; # Report the DRC Setup
    signoff_drc –check_all_layers –run_dir {$result_dir} ; # DRC Setup
    process_drc_report –mode drc –run_dir $result_dir;
}

# DRC+ Physical Verification
if {[file exists $drcplus_runset]} {
    set_physical_signoff_options –exec_cmd {icv} –drc_runset {$drc_runset}; # DRC Setup
    report_physical_signoff_options; # Report the DRC+ Setup
    signoff_drc –check_all_layers –run_dir {$result_dir} ; # DRC+ Setup
    process_drc_report –mode drcplus –run_dir $result_dir;
}
```

## III.  Result

In this work, we have evaluated seven 14-nm standard cells libraries to validate our proposed Synopsys ICC/ICV-based integrated utility tool. The enumerated layouts are verified with our Synopsys ICV DRC kit using the ICC In-Design physical verification feature to accelerate design rule convergence.

During the layout stream-out phase with the Synopsys ICC, designers have the options either to use the pre-defined coloured standard cell layout (DPT Option I) or to perform DPT using the Synopsys ICV DRC kit (DPT Option II).  These two options were evaluated through our utility and the DRC and DRC+ verification results are presented in the Table 2.

From our experimental result with DPT Option I, we have observed that there are three out of seven libraries, having DRC+ violations. On the other hand, there is no DRC and DRC+ violation for DPT option II. The experimental results match well with our expectations. This is because DPT Option II will be able to handles all the cell boundary conditions related to color decomposition.
Designers have to be aware that standard cells' spice models are highly dependent on the layout coloring. Even though DPT Option II is able to avoid any coloring violation, it might affect the circuit performance. Therefore, it is highly advisable for IP designers to fix the entire DRC+ violations before releasing the IP to their customers. The IP designers can use our utility to identify all the possible abutment conditions.

Table 2: Coloring-related DRC and DRC+ violations of standard cells.

| Libraries | DPT Option I | | DPT Option II | |
|---|---|---|---|---|
| | DRC error count | DRC+ error count | DRC error count | DRC+ error count |
| 1 | Clean | **218** | Clean | Clean |
| 2 | Clean | **196** | Clean | Clean |
| 3 | Clean | Clean | Clean | Clean |
| 4 | Clean | Clean | Clean | Clean |
| 5 | Clean | **815** | Clean | Clean |
| 6 | Clean | Clean | Clean | Clean |
| 7 | Clean | Clean | Clean | Clean |

The most common DPT coloring-related DRC and DRC+ violations is the metal line bridging due to insufficient spacing.  Figure 12 describes the DRC+ hotspot condition where two M1_E1 and two M1_E2 metal lines with certain run length aligned in parallel with certain spacing. The metal line bridging occurs between the M1_E1 and M1_E2 metal lines. For illustration of the DRC+ violation, we have provided the corresponding violation location in the layout as shown in Figure 14. This violation is observed using the DPT Option I. However, using the DPT Option II, the DRC-based DPT has corrected the coloring problem in the layout as shown in Figure 13.

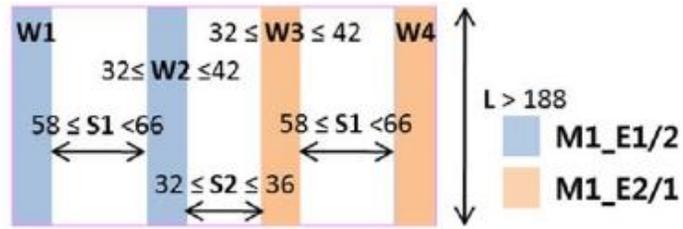

Figure 12: DRC+ hotspot pattern.

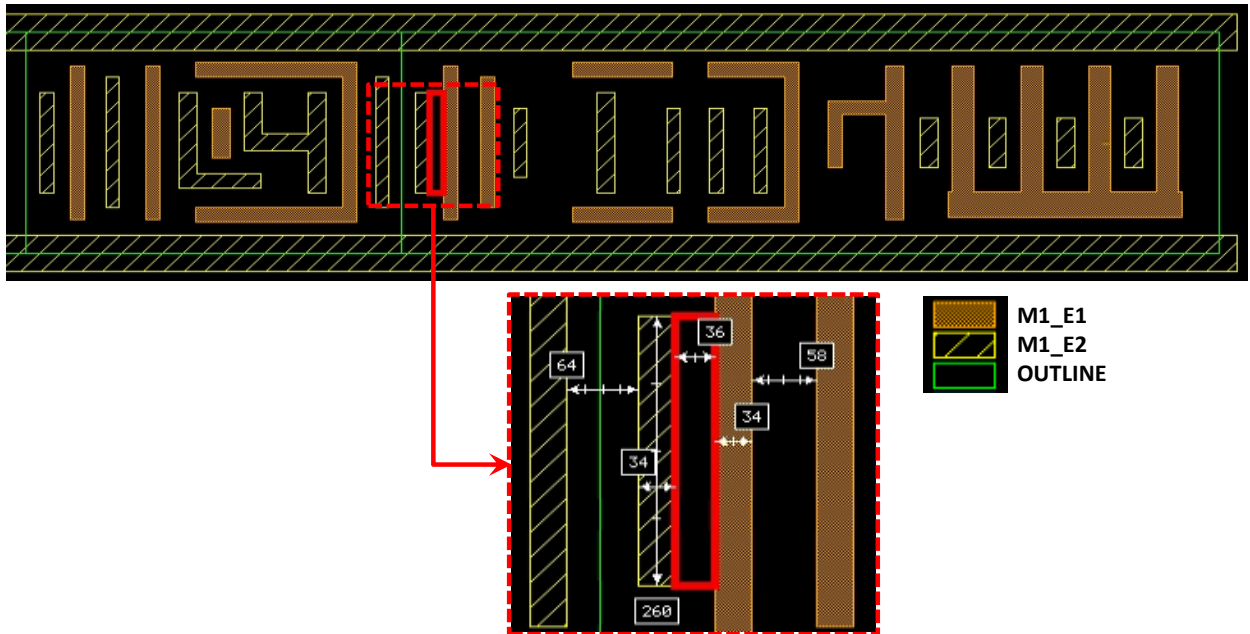

Figure 13: DRC+ hotspot found at boundary for standard cells that are fixed coloring before abutment placement (DPT Option I).

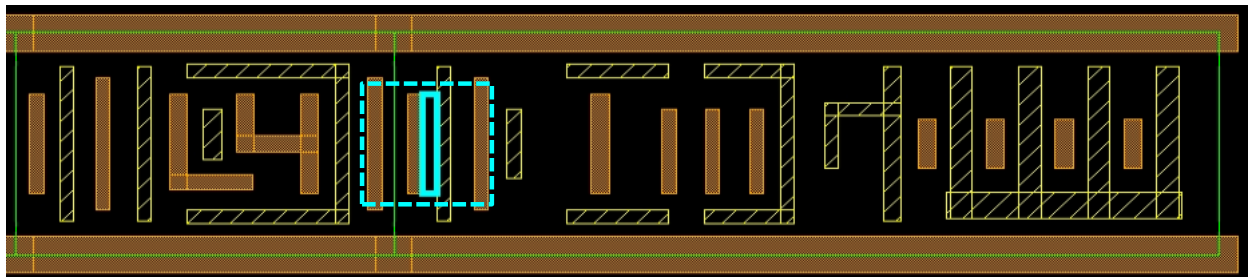

Figure 14: No coloured-related violation for standard cells that are colored only after abutment placement (DPT Option II).

## IV. Conclusions

In this work, we have developed an easy-to-use utility for library physical layout designer to validate the design- and lithography-compliances of their standard cell library through the Synopsys IC Compiler. To achieve 100% standard cell abutment coverage with minimum verification runtime, we have proposed an area-efficient abutment placement, which achieves a 3.2 times reduction compared to the conventional single-height side-side standard cell abutments. This significantly reduces the runtime, memory and minimizes redundant violation errors. The proposed utility has been validated with several standard cell libraries across multiple technologies.